# Two Issues in Modelling Fish Migration[1]

Hidekazu Yoshioka
*Graduate School of Advanced Science and Technology*
*Japan Advanced Institute of Science and Technology*
Nomi, Japan
yoshih@jaist.ac.jp

*Abstract*—Fish migration is a dynamic phenomenon observed in many surface water bodies on the earth, while its understanding is still insufficient. Particularly, the biological mechanism behind fish migration is not fully understood. Moreover, its observation is often conducted visually and hence manually, raising questions of accuracy and interpretation of the data sampled. We address the two issues, mechanism and observation, of fish migration based on a recently developed mathematical model. The results obtained in this short paper show that fish migration can be characterized through a minimization principle and evaluate the error of its manual observations. The minimization principle we hypothesize is an optimal control problem where the migrating fish population dynamically changes its size and fluctuation. We numerically investigate alternating and intensive observation schemes as case studies, demonstrating that in some realistic conditions the estimate of total fish count is not reliable. We believe that this paper contributes to a deeper understanding of fish migration.

## I. Introduction

Sustainable food production is an indispensable task for human lives. Its achievement needs proper understanding about food resources. In fisheries management in particular, understanding about behavioral characteristics of fish species is crucial because they often migrate a long distance depending on environmental and social cues; even freshwater fish species migrate several tens kilometers for spawning [1]. It means that fish population would vary both in space and time, and their evaluation is crucial for effective fisheries management. Moreover, fish migration is a driver of ecosystem services, and hence its analysis is also important from an ecological viewpoint [2].

Fish migration is a dynamic biological phenomenon that often occurs seasonally to complete a lifetime event such as spawning [3-5], while it has recently been found that migration has a multiscale nature at least in time. Yoshioka [6] analyzed the sub-hourly migration data of the diadromous fish (i.e., it migrates between river and sea) *Plecoglossus altivelis altivelis* in Japan, commonly called *Ayu* in the country. The fish has a one-year life cycle, and the study [6] focused on its upstream migration from a sea to a river during the juvenile stage (**Fig. 1**). The data, which were acquired through an AI-assisted video monitoring system, counted individual fishes passing through a fixed observation point of a river in each 10-min interval during daytime for four months in 2023 and 2024. An important finding was that the fish count significantly fluctuates within a day, showing an intermittency with an on-off behavior [6] (**Fig. 2**). However, the biological mechanism behind this phenomenon has not been clarified.

So, why is such a fine analysis important in practice? One motivation behind the previous study [6] was that counting *Ayu* during its upstream migration stage, which can be important for assessing its population dynamics, has mainly been conducted manually. Manually (and visually) counting fish migrating along a river would be a demanding work; for *Ayu*, it is often the case that more than millions of individuals migrate each year along a river [7-9]. Usually, manually counting the migrating fish is not fully covering an entire day, e.g., alternate 15-min count and 15-min rest [9] (called *alternating observation* in this paper) or count during a certain time interval (less than one to several hours) in a day (called *intensive observation* in this paper) [10-12]; some intensive observation manually counts fish in an underwater video footage. The intermittency observed in the 10-min migrating fish count data [6] implies that the manual fish count may be polluted by errors if burst and rest events are not captured during observations. Indeed, there is a report that the manual observation underestimates the video one by about 60% [11]. These findings motivate us to investigate errors introduced by partial observations of fish migration.

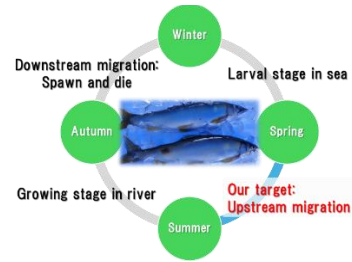

Fig. 1. Life cycle of the fish *Ayu*. This paper focuses on juvenile upstream migration.

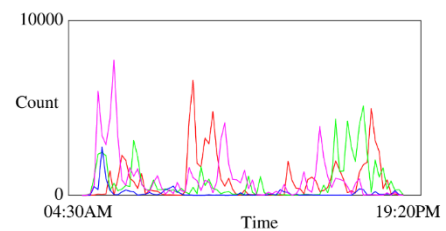

Fig. 2. Examples of 10-min time series data of *Ayu* observed at the Nagara River Barrage in 2024: April 26 (red), April 27 (green), April 28 (blue), and April 29 (magenta). Intermittency in data is visible.

Motivated by the issues of the mechanism and observation scheme of fish migration, the aim of this short paper is to address them based on the validated mathematical model of Yoshioka [6] tailored for the juvenile upstream migration of *Ayu*. The model is a stochastic differential equation (SDE)[13] having a blowing-up coefficient, with which the diurnal migration behavior as well as intermittency can be studied.

---

We show that the coefficient can be understood as the optimal control of a stochastic control problem representing an energy minimization principle. This theoretically characterizes fish migration as a solution to a minimization problem under certain limited conditions, which is a novel viewpoint for understanding this complex phenomenon. Then, we address the issue of partial observation by a Monte-Carlo method for simulating alternating and intensive observation schemes.

## II. THE SDE

### A. Formulation

In this paper, we do not review SDEs and their optimal control in general; interested readers can read the textbook [13]. All SDEs appearing in this paper are considered in the Itô's sense. Let $t \geq 0$ be time and $W_t$ be a 1-D standard Brownian motion at time $t$. Let $X_t$ be the unit-time fish count of juvenile *Ayu* at a fixed point river at time $t$, which is a non-negative and continuous-time random variable.

We consider the fish migration within one day, which is represented by a time interval $[0,T]$, where the time 0 and $T > 0$ correspond to the sunrise time and sunset time, respectively. Particularly, the upstream migration of *Ayu* has been suggested to occur only during daytime, and hence $X_0 = X_T = 0$ is assumed. With these preparations, the SDE that governs $X_t$ ($0 < t < T$) was proposed heuristically as follows [6] (called a diffusion bridge in this paper):

$$\mathrm{d}X_t = \underbrace{(a - h_t X_t)\mathrm{d}t}_{\text{Deterministic component}} + \underbrace{\sigma\sqrt{h_t X_t}\mathrm{d}W_t}_{\text{Stochastic component}}, \quad 0 < t < T. \quad (1)$$

Here, $a$ (rate of increase of the fish count) and $\sigma$ (intensity of fluctuation in the fish count) are positive parameters, and $h$ (scaling function that controls the mean reversion and fluctuation) is a positive function of $t$ that is continuously differentiable for $0 < t < T$ and blows up to $+\infty$ as $t \to T$.

The fitting results [6] suggested that $\sigma^2$ is significantly larger than $a$, generating highly fluctuating and intermittent behavior of $X_t$ as for the empirical data in **Fig. 2**. See **Fig. 3**.

### B. On the Unbounded Coefficient

We summarize the discussion in Yoshioka [6] along with a couple of remarks. The SDE admits a unique pathwise continuous solution if $h$ blows up as $t \to T$ with the speed of $(T - t)^{-1}$. From a mathematical standpoint, this particular speed of blowing up of $h$ guarantees the unique existence of solutions to the SDE (1) that satisfies the terminal condition $X_T = 0$ with probability 1. The coefficient $h$ appears in both drift and diffusion terms of the SDE (1), and it has the meaning of the speed of increase of the modified time, i.e., a biological clock of fish migration, that is considered to implicitly reflect the environmental cues such as diurnal changes of water temperature and irradiation.

Clearly, the coefficient $h$ is a key element in the SDE (1). So, theoretically where does it come from? We try to answer to this question in the next section. We show that $h$ of the following form arises from a minimization principle:

$$h_t = \frac{c}{T-t}, \quad 0 < t < T. \quad (2)$$

Here, $c$ is a positive constant. The coefficient $h$ in (2) corresponds to "model 1" [6] that has been applied to real data.

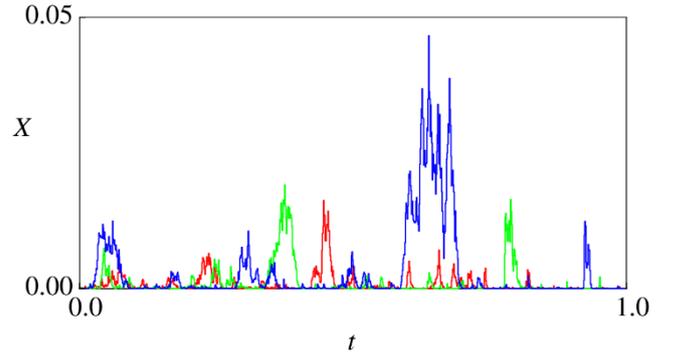

Fig. 3. Sample paths of "model 1" [6] in 2024. Here, we set the terminal time to be 1. Colors represent distinct sample paths. Intermittency of sample paths is visible.

## III. VARIATIONAL CHARACTERIZATION

### A. Formulation

We formulate a stochastic control problem of minimizing an expectation that represents some energy needed to complete the upstream migration in one day. Namely, we hypothesize that migration results from a minimization principle of energy. Our control formalism basically follows that of the common dynamic programming principle (e.g., Chapter 5 [13]).

We consider the following SDE

$$\mathrm{d}X_t = (a - u_t X_t)\mathrm{d}t + \sigma\sqrt{u_t X_t}\mathrm{d}W_t, \quad 0 < t < T \quad (3)$$

with $X_0 = X_T = 0$, which is analogous to (1) but the coefficient $h$ is now replaced by a non-negative control variable $u$ to be optimized so that the following conditional expectation $J$ is minimized for each $t \in (0,T)$ and $x > 0$:

$$J(t,x;u) = E\left[\int_t^T \frac{1}{c} X_s u_s^c \mathrm{d}s \bigg| X_t = x\right] \quad (4)$$

along with the constraint of the control $u_t$ ($0 < t < T$); $X$ solves the SDE (3) and satisfies $X_0 = X_T = 0$. Here, $E$ is the expectation and $c \in (1,2)$ is a constant. The meaning of the range of the parameter $c$ will be clarified later.

The right-hand side of (4) represents the cumulated unit-time energy (the integrand) required for completing the migration during $(t,T)$ under the condition $X_t = x$. The decision-maker of the control problem here is the migrating fish population. The functional form of the unit-time energy is motivated from the ansatz that the energy needed for the migration is proportional to the migrating population and that specifying a larger value of the control variable would incur a larger energetic cost.

The value function as the minimized $J$ is defined as

$$\Phi(t,x) = \inf_u J(t,x;u). \quad (5)$$

Recall that the control $u$ needs to be chosen so that $X_0 = X_T = 0$, and it should be non-negative. Actually, there exists technical conditions such as measurability [13], but is not detailed in this paper. The minimizing $u$ characterizes the fish migration as verified in the next sub-section.

*B. Solution*

We want to solve (5); however, directly dealing with it is difficult due to the prescribed terminal condition $X_T = 0$. We follow the methodology of Yoshioka and Yamazaki [14] that approximates a control problem with a terminal constraint with a penalized version without the constraint. The penalization problem in our case is given as follows:

$$\Phi_\eta(t,x) = \inf_u J_\eta(t,x;u) \tag{6}$$

with

$$J_\eta(t,x;u) = E\left[\int_t^T \frac{1}{c} X_s u_s^c \, \mathrm{d}s + \eta X_T \,\bigg|\, X_t = x\right], \tag{7}$$

where $\eta > 0$ is a penalization parameter. In (6)-(7), controls $u$ are non-negative and are free from the constraint $X_T = 0$. The added term $\eta X_T$ in (7) increases as $X_T$ deviates from 0, and the degree of its penalization is modulated by $\eta$.

The solution strategy that we chose here is to firstly solve the control problem with penalization (6) and then take the limit $\eta \to +\infty$ at which the terminal condition $X_T = 0$ would be satisfied with probability 1; otherwise, the right-hand side of (7) possibly diverges. The dynamic programming principle [13] suggests that $\Phi_\eta$ is governed by the nonlinear partial differential equation (PDE):

$$\frac{\partial \Phi_\eta}{\partial t} + a\frac{\partial \Phi_\eta}{\partial x} + x\inf_{w \geq 0}\left\{-w\frac{\partial \Phi_\eta}{\partial x} + \frac{\sigma^2}{2}w\frac{\partial^2 \Phi_\eta}{\partial x^2} + \frac{w^c}{c}\right\} = 0 \tag{8}$$

for $x \geq 0$ and $0 < t < T$ with the terminal condition $\Phi_\eta(x,T) = \eta x$ for $x \geq 0$.

The verification argument [13] shows that the following solution to the PDE (8) exists:

$$\Phi_\eta(t,x) = A_{\eta,t} x + B_{\eta,t} \tag{9}$$

for $x \geq 0$ and $0 < t < T$ with ordinary differential equations

$$\frac{\mathrm{d}A_{\eta,t}}{\mathrm{d}t} = \frac{c-1}{c}\left(A_{\eta,t}\right)^{\frac{c}{c-1}}, \ 0 < t < T, \tag{10}$$

$$\frac{\mathrm{d}B_{\eta,t}}{\mathrm{d}t} = -aA_{\eta,t}, \ 0 < t < T \tag{11}$$

subject to $A_{\eta,T} = \eta$ and $B_{\eta,T} = 0$. Moreover, we obtain that the optimal control $u = u_\eta^*$ as the minimizer in (6) is given by

$$u_{\eta,t}^* = \left(A_{\eta,t}\right)^{\frac{1}{c-1}}, \ 0 < t < T. \tag{12}$$

We can exactly solve (10) as follows:

$$A_{\eta,t} = \left(\eta^{\frac{-1}{c-1}} + \frac{1}{c}(T-t)\right)^{-(c-1)}, \ 0 < t \leq T. \tag{13}$$

Then, by (12), we arrive at the following optimal control of the penalized problem:

$$u_{\eta,t}^* = \left(\eta^{\frac{-1}{c-1}} + \frac{1}{c}(T-t)\right)^{-1}, \ 0 < t < T. \tag{14}$$

Because $c > 1$, sending $\eta \to +\infty$ in (14) formally yields (2).

Now, we could obtain that the coefficient $h$ can be characterized as the optimal control. This is a new characterization of the migration of *Ayu*. Particularly, the speed of blow up $(T-t)^{-1}$ near the terminal time $T$ is optimal in view of the minimization problem. However, there is a serious limitation of the theoretical result obtained in this sub-section: the condition $c \in (1,2)$ of the coefficient. The lower-bound $c > 1$ to obtain a bounded optimal control when $\eta \to +\infty$; see the right-hand side of (14). The upper bound $c < 2$ is necessary to guarantee the boundedness of the right-hand side of (4) with $u = u^*$ due to

$$0 \leq J(0,x;u^*) = \frac{ac^{c-1}}{2-c}, \tag{15}$$

where the condition $c < 2$ was necessary to derive the right-most result of (15), and $E[X_s]$ is obtained by directly taking the expectation of both sides of the SDE (3). Moreover, when $c \in (1,2)$, an elementary calculation gives $E[\eta X_T] \to 0$ as $\eta \to +\infty$, while the limit is a bounded positive value when $c = 2$ and it diverges to $+\infty$ when $c > 2$. Therefore, the condition $c \in (1,2)$ is needed to obtain the convergence the optimal control of the penalized problem to that of the original one. If this consistency condition is abandoned, then we can formally allow for $c \geq 2$ but then $\Phi_\eta$ diverges to $+\infty$ under the limit $\eta \to +\infty$ as explained above.

According to the fitted model against real data [6], fitted values of $c$ are not necessarily between 1 and 2. Despite the limitations discussed above, the theoretical results obtained here partly justifies the modelling strategy employed in Yoshioka [6] for the first time. In another work, we are currently investigating cases where $c$ does not fall on this range, by using a modified minimization principle.

## IV. PARTIAL OBSERVATION

*A. Problem Setting*

We turn to the issue of partial observation schemes. The focus in **Section IV** is the difference between the true total fish count and partially observed ones in one day. Without significant loss of generality, we set the non-dimensionalization $T = 1$, and hence the sunrise time is 0 and sunset time is 1. By a true observation $O$, we mean

$$O = \int_0^1 X_s \, \mathrm{d}s \tag{16}$$

because $X_t$ is the unit-time fish count at time $t \in (0,1)$.

We consider the two partial observation schemes: alternating observation and intensive observation. The estimated fish count by the alternating observation is

$$O_{\text{Alt}} = 2 \sum_{i=\text{odd}}^{N} \int_{(i-1)d}^{id} X_s \, \mathrm{d}s, \tag{17}$$

where $N$ is an even natural number and $d = N^{-1}$. In the alternating observation, counting fish is conducted during the time intervals $(0, d)$, $(2d, 3d)$, $(4d, 5d)$ …, and the total observation duration always equals 0.5, which is why "2" is multiplied in the right-hand side of (17). In the alternating observation, $d$ has been chosen to be 10 to 15 min, e.g., [9].

The estimated fish count by the intensive observation is

$$O_{\text{Int}} = \frac{1}{l} \int_{\tau - \frac{l}{2}}^{\tau + \frac{l}{2}} X_s \, \mathrm{d}s, \tag{18}$$

where $\tau \in (0, 1)$ and $l > 0$ is the length of observation chosen so that $0 \leq \tau - l/2 \leq \tau + l/2 \leq 1$. In the intensive observation, individual fishes are counted only around the time $\tau$ with the observation duration $l$, which is why "$l$" is multiplied in the right-hand side of (18). In the intensive observation, $l$ is typically set as few hours, e.g., [10,11].

The relative error between the true and partial observations is defined as follows:

$$r = \frac{O_Z}{O} - 1, \quad Z = \text{Alt or Int}. \tag{19}$$

There is no relative error if $r = 0$, and there is underestimation and overestimation if $r < 0$ and $r > 0$, respectively. The minimum value of the relative error is $-1$, while its maximum value is unbounded. Note that $O$ and $O_Z$ are correlated random variables for each sample path of $X$, and we do not know the closed-form statistics of the relative error $r$. Therefore, we compute the relative error $r$ using a Monte-Carlo method [6] with the time increment of 1/100,000 and the number of sample paths of 1,000,000.

As the SDE for *Ayu*, we use the following normalized "model 2" in 2024 [6] that better fits than "model 1":

$$h_t = \frac{1}{t + \varepsilon} + \frac{1}{1 - t}, \quad 0 < t < 1 \tag{20}$$

with $\varepsilon > 0$. The length of time interval $[0, 1]$ corresponds to 11 to 15 hours. We fix $\tau = 0.5$, which is around the daytime.

### B. Results and Discussion

**Table I** shows the computed average, standard deviation, and skewness for the relative error $r$ of the alternating observations. Similarly, **Table II** shows those of the intensive observations. **Tables III-IV** show the corresponding results for the absolute value $|r|$. **Figs. 4-5** show the computed probability density functions (PDFs) of relative error $r$ for alternating and intensive observations, respectively.

For the alternating observations, the average of the relative error $r$ can be made very small (less than 0.1%) even for a relatively large $d$ such as 0.25; however, the standard deviation becomes around 10% to 50% in the computed cases, and the PDFs of the relative error becomes more uniform between $-1$ and 1 as $d$ increases (**Fig. 4** and **Table I**). The uniform-like distribution implies that the observation is highly randomized and hence is not reliable at all. This point is quantified in **Table III** for $|r|$. The choice of error metric is therefore critical in this case. For the finest alternating observation examined here ($d = 0.01$) corresponds to around 10 minutes; in this fine-observation case, the standard deviation of relative error can be made less than 10%.

TABLE I. AVERAGE, STANDARD DEVIATION, AND SKEWNESS OF THE ALTERNATING OBSERVATIONS OF RELATIVE ERROR

| $d$ | Average | Standard deviation | Skewness |
|---|---|---|---|
| 0.010 | -5.48.E-05 | 8.11.E-02 | -1.03.E-02 |
| 0.025 | -1.80.E-04 | 1.75.E-01 | -5.46.E-03 |
| 0.050 | -3.10.E-04 | 2.81.E-01 | -1.48.E-03 |
| 0.100 | -4.87.E-05 | 3.98.E-01 | 1.38.E-03 |
| 0.250 | -6.20.E-04 | 5.32.E-01 | 6.41.E-04 |

TABLE II. AVERAGE, STANDARD DEVIATION, AND SKEWNESS OF THE INTENSIVE OBSERVATIONS OF RELATIVE ERROR

| $l$ | Average | Standard deviation | Skewness |
|---|---|---|---|
| 0.20 | 2.37.E-01 | 1.16.E+00 | 1.02.E+00 |
| 0.50 | 1.96.E-01 | 5.60.E-01 | -3.90.E-01 |
| 0.60 | 1.74.E-01 | 4.32.E-01 | -8.58.E-01 |
| 0.80 | 1.11.E-01 | 2.11.E-01 | -2.31.E+00 |
| 0.90 | 6.64.E-02 | 1.07.E-01 | -4.25.E+00 |

TABLE III. AVERAGE, STANDARD DEVIATION, AND SKEWNESS OF THE ALTERNATING OBSERVATIONS OF ABSOLUTE RELATIVE ERROR

| $d$ | Average | Standard deviation | Skewness |
|---|---|---|---|
| 0.010 | 5.55.E-02 | 5.91.E-02 | 2.23.E+00 |
| 0.025 | 1.27.E-01 | 1.21.E-01 | 1.56.E+00 |
| 0.050 | 2.16.E-01 | 1.80.E-01 | 1.02.E+00 |
| 0.100 | 3.25.E-01 | 2.31.E-01 | 5.55.E-01 |
| 0.250 | 4.57.E-01 | 2.72.E-01 | 9.18.E-02 |

TABLE IV. AVERAGE, STANDARD DEVIATION, AND SKEWNESS OF THE INTENSIVE OBSERVATIONS OF ABSOLUTE RELATIVE ERROR

| $l$ | Average | Standard deviation | Skewness |
|---|---|---|---|
| 0.20 | 9.13.E-01 | 7.47.E-01 | 1.55.E+00 |
| 0.50 | 5.20.E-01 | 2.86.E-01 | -1.20.E-01 |
| 0.60 | 4.14.E-01 | 2.14.E-01 | -1.20.E-01 |
| 0.80 | 2.09.E-01 | 1.15.E-01 | 2.17.E+00 |
| 0.90 | 1.05.E-01 | 6.87.E-02 | 5.45.E+00 |

For the intensive observations, both the average and standard deviation of relative error is about 10% even if one covers 90% ($l = 0.90$) of the whole daytime (time interval $(0, 1)$) as shown in **Table II**. The relative error is on average 24% with the standard deviation more than 100% if the observation is made with the duration $l = 0.20$ corresponding to 2 to 3 hours in the real world. Parameter dependence of the skewness of $|r|$ is monotone for the alternating observation, but is less clear and non-monotone for the intensive observation. The size of relative error decreases as the duration $l$ increases in the intensive observation as visualized in **Fig. 5**. Comparing **Figs. 4-5** show that the relative error distributions are close to be symmetric for the alternating observation while they are possibly visibly skewed for the intensive observation. Hence, they are qualitatively different in terms of relative errors. Comparing **Table II** for $r$ and **Table IV** for $|r|$ suggest that both error metrics perform qualitatively the same for the intensive observation.

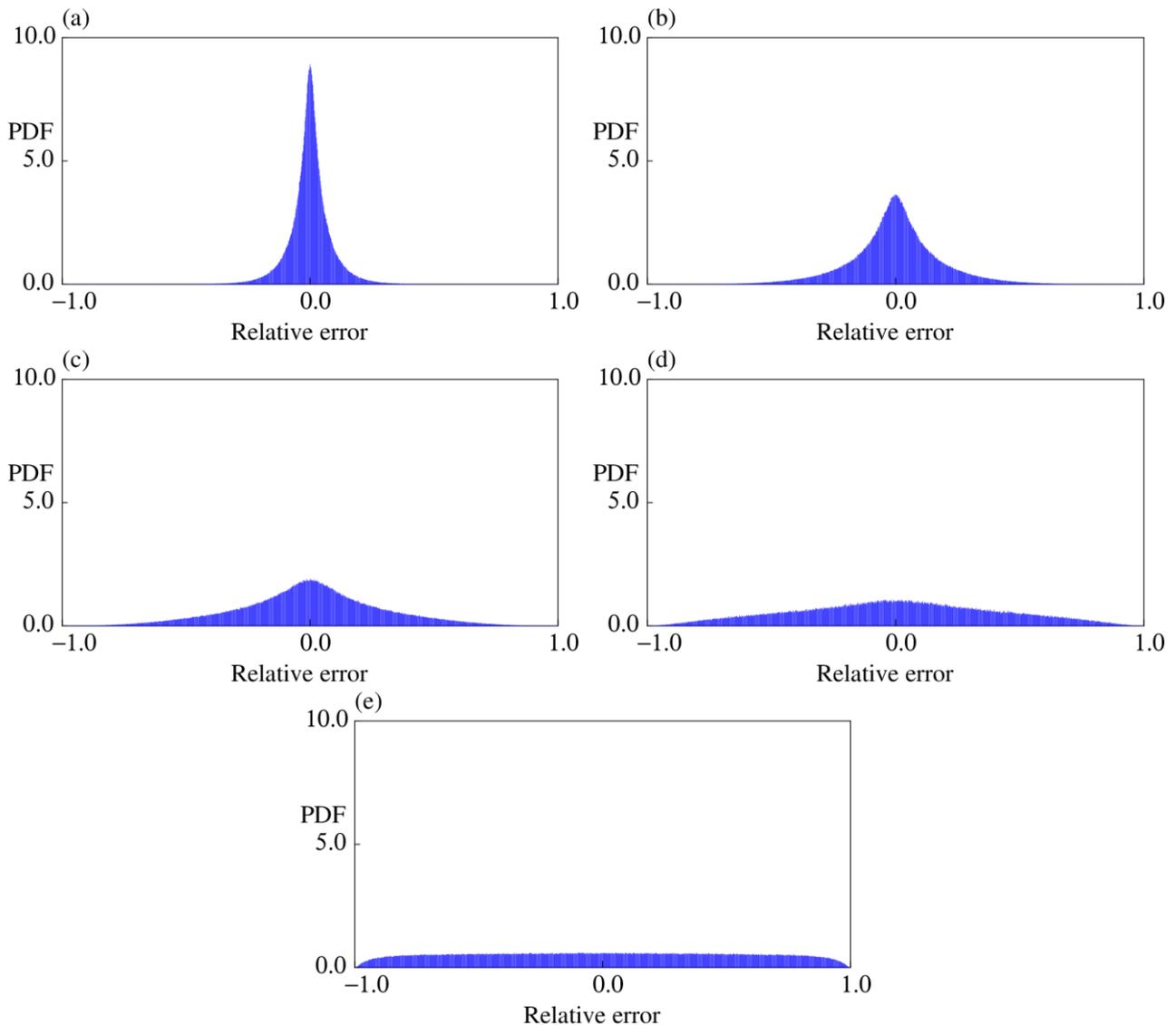

Fig. 4. Computed histograms of relative error in the alternating observation. Values of *d* are: (a) 0.010, (b) 0.025, (c) 0.050, (d) 0.100, and (e) 0.250.

The computational results suggest that interpretation of the existing manually-obtained data should be with care. We did not account for human errors; therefore, the actual accuracy of the existing data may be worse than that we predicted. The result obtained here is thus rather optimistic; adding human error would increase the variance of relative error. It is also important to investigate how the target data of fish migration were acquired. Another factor that we did not consider is labor costs of observations. Labor costs would be proportional to the duration of observation; one may be able to design a cost-efficient observation scheme based on our results. Our results suggest that the alternating observation outperforms the intensive ones for each fixed total observation duration (0.5 in this paper) if $d$ is small. This is considered due to that the alternating scheme is able to track fluctuation of intermittent sample paths if $d$ is small. However, ease of implementation of each observation scheme may be case sensitive.

## V. Conclusion

We briefly reviewed an SDE model that has recently been proposed for modelling the upstream migration of *Ayu*. The first part of this short paper showed that the model can be obtained through a minimization principle of a stochastic control problem, adding a new theoretical characterization to the heuristics made in the paper [6]. The second part of this paper addressed partial observation problems of the upstream migration of *Ayu*. We numerically computed errors induced by the two conventional manual observation schemes, evaluating the relationship between observation schemes and their errors. Our findings suggest that accuracy of existing fish count data obtained manually should be (re)considered by how they were acquired.

We are still on the way. Finding a clear and beautiful law behind each complex phenomenon is the absolute task of researchers forever. At least, the author believes so.

## Acknowledgment

This study was supported by the Japan Science and Technology Agency (PRESTO No. JPMJPR24KE). The author thanks Ibi River and Nagara River General Management Office for providing valuable data about upstream migration of *Ayu*.

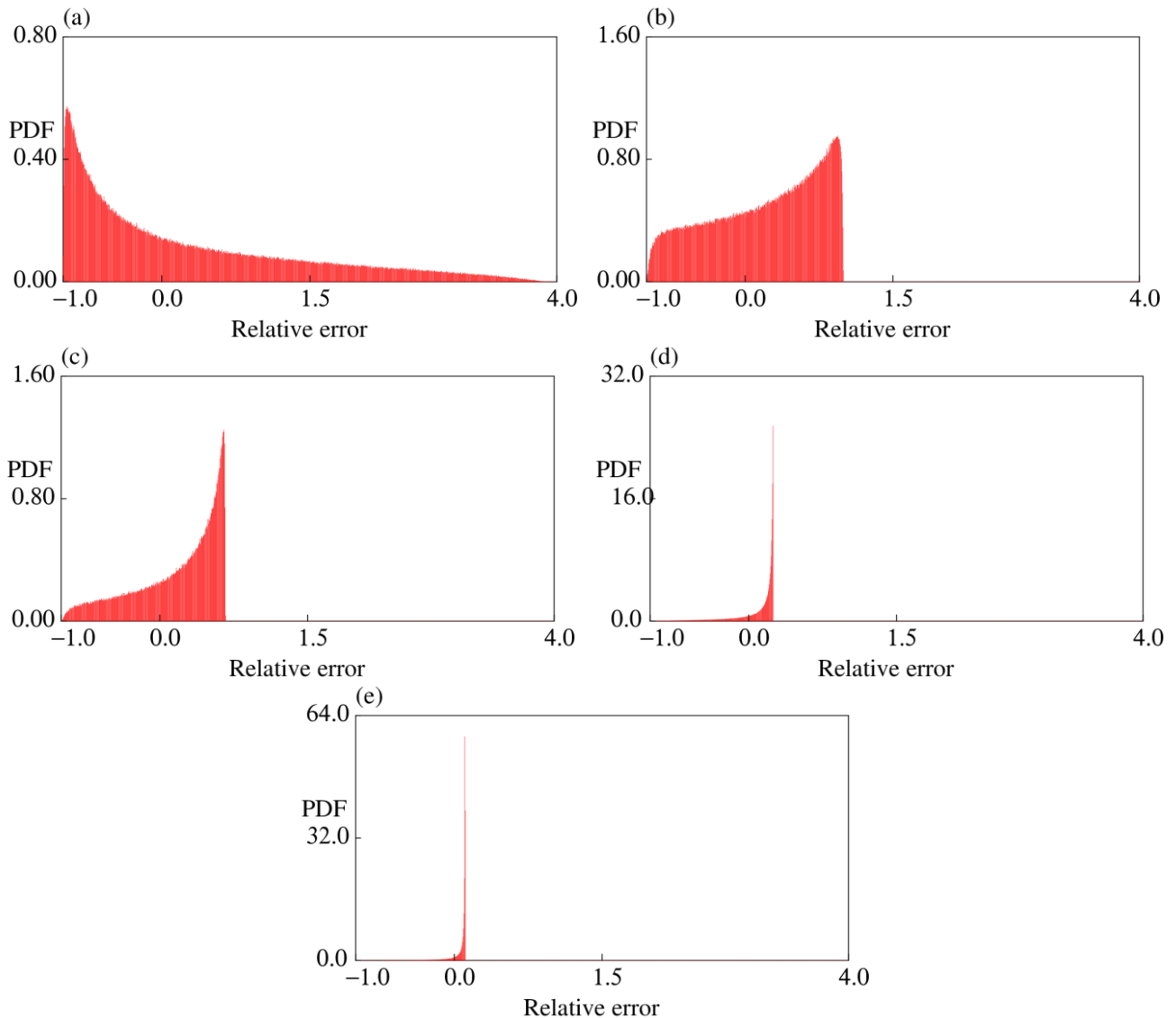

Fig. 5. Computed histograms of relative error in the intensive observation. Values of $l$ are: (a) 0.20, (b) 0.25, (c) 0.60, (d) 0.80, and (e) 0.90.